\documentclass[aps,preprint,amsmath,amsfonts,amssymb,showpacs,showkeys]{revtex4}
\usepackage{bm}
\usepackage{epsfig}
\newcommand{\x}{{\bf{x}}}
\newcommand{\y}{{\bf{y}}}

\newcommand{\act}{{\sf act}}
\newcommand{\lis}{{\sf list}}
\newcommand{\brh}{{\bm{\rho}}}
\begin{document}
\title{Monte Carlo simulations of bosonic reaction-diffusion systems and comparison to Langevin equation description}
\author{Su-Chan Park}
\email{psc@kias.re.kr}
\affiliation{Korea Institute for Advanced Study, Seoul 130-722, Korea}
\date{\today}
\begin{abstract}
Using the Monte Carlo simulation method for bosonic 
reaction-diffusion systems introduced recently [S.-C. Park,
Phys. Rev. E {\bf 72}, 036111 (2005)], one dimensional bosonic  models
are studied and compared to the corresponding Langevin equations 
derived from the coherent state path integral formalism.
For the single species annihilation model, the exact asymptotic form
of the correlation functions is conjectured and the full equivalence
of the (discrete variable) master equation and the (continuous variable)
Langevin equation is confirmed numerically.
We also investigate the cyclically coupled model of bosons which 
is related to the pair contact process with diffusion (PCPD).
From the path integral formalism, Langevin equations which are expected to
describe the critical behavior of the PCPD are
derived and compared to the Monte Carlo simulations of the discrete model.
\end{abstract}
\maketitle
\section{Introduction}
The reaction-diffusion (RD) systems have played a paradigmatic role in studying
certain physical, chemical, and biological systems \cite{P97}.
In the study of the RD systems on a lattice via Monte Carlo (MC) simulations,
particles are usually assigned hard core exclusion property.
On the other hand, the renormalization group (RG) calculations which 
have been successfully applied to several RD systems
are often performed with boson systems \cite{bosonF,Lee94,CT96}.
Hence, the comparison of the numerical studies 
with the RG calculations can sometimes become a nontrivial issue.

There are two ways to bridge this gap between numerical and analytical studies.
One is to make a path integral formula for hard core particles 
which is suitable for the RG calculations.  This path has indeed been 
sought and some formalisms are suggested \cite{PKP00,vW01,PjP05}.
The other is to find a numerical method that would simulate boson systems.
In this context, numerical integration studies of equivalent 
Langevin equations to the boson systems have been performed, too 
\cite{BAF97,CD99,PL99,DCM04}.
However, it is not always possible to find an equivalent Langevin equation
\cite{G83} and hence the applicability of this approach is 
somewhat restricted. Therefore, another numerical method is called for.

Recently, a general algorithm to simulate the bosonic RD systems was
proposed \cite{P05c}.
Section \ref{Sec:algo} is devoted to a heuristic explanation of 
this algorithm to simulate general bosonic RD systems.
In Sec. \ref{Sec:app}, the numerical method is applied to two
bosonic RD systems. First, the single species annihilation
model is studied with the emphasis on the pair correlation functions.
We conjecture an exact asymptotic behavior of these quantities. 
We then present the numerical comparison of the discrete model to  
Langevin equation of continuous variables.
Then, the cyclically coupled  model of bosons is introduced and 
Langevin equations for this model with/without
bias are derived from the well-trodden path integral formalism and 
compared to MC simulations.
Section \ref{Sec:sum} summarizes the work.

\section{\label{Sec:algo}Algorithm}
This section explains the method proposed in Ref. \cite{P05c} that is 
suitable for MC simulations of bosonic RD systems.  After describing how 
single species boson systems can be simulated, a brief remark regarding 
the generalization to multiple species will be followed.

The reaction dynamics of diffusing bosons is represented as
\begin{equation}
n A \stackrel{\lambda_{nm}}{\longrightarrow}  (n+m) A,
\label{Eq:dynamics}
\end{equation}
where $n\ge 0$, $m \ge -n$ ($m\neq 0$),
 and $\lambda_{nm}$ is the transition rate.
Each particle diffuses with rate $D$ on a $d$ dimensional hypercubic lattice.
The periodic boundary conditions are always assumed, but 
other boundary conditions do not limit the validity of the algorithm below.
Configurations are specified by the occupation number 
$\rho_\x$ ($\ge 0$) at each lattice point $\x$. 
A configuration is denoted as $\{ \rho \}$
which means $\{\rho_{\x} | \x \in {\bf L}^d\}$,
where ${\bf L}^d$ stands for the set of lattice points and the cardinality
of ${\bf L}^d$ is $L^d$.  

The master equation which describes 
stochastic processes modeled by Eq. (\ref{Eq:dynamics}) takes the form
\cite{G83,K97}
\begin{equation}
\begin{aligned}
&\frac{\partial P }{\partial t} = D \sum_{\langle \x ,\y \rangle} 
\left ( (\rho_\x + 1)  \hat E_{\x,\y} - \rho_\x \right )P\\
&+ \sum_{n,m} \lambda_{nm} \sum_{\x} \left ( f(\rho_\x-m,n) \hat C_{\x,m} 
 - f(\rho_\x ,n)  \right )P,
\end{aligned}
\label{Eq:master}
\end{equation}
where $P=P(\{\rho\},t)$ is the probability with which the configuration
of the system is $\{\rho\}$ at time $t$, 
$\langle \x,\y \rangle$ means the nearest neighbor pair ($\x,\y \in {\bf L}^d$), 
$f(\rho_\x,n) = (\rho_\x !)/(\rho_\x -n )!$ is the number of
ordered $n$-tuples at site $\x$ of the configuration $\{\rho\}$,
and $\hat E_{\x,\y}$ and $\hat C_{\x,m}$ are operators affecting
$P(\{\rho\},t)$ such that
\begin{equation}
\begin{aligned}
\hat E_{\x,\y} P  &= P(\{\cdots,\rho_\x +1,
\rho_\y-1,\cdots\};t),  \\
\hat C_{\x,m}P  &= P(\{\cdots,\rho_\x-m,\cdots\};t). 
\end{aligned}
\end{equation}

The master equation implies that the average number of transition events 
for the configuration $\{\rho\}$ during infinitesimal time interval $dt$  is
\begin{equation}
\label{Eq:ave}
\begin{aligned}
E(dt,&\{\rho\})=
dt \sum_{\x,n} \left ( 2 d D\delta_{n,1} + \sum_m \lambda_{nm} 
\right ) f(\rho_\x,n)\\
&=dt \sum_{\x,n} \left ( 2 d D\delta_{n,1} +  \sum_m n!\lambda_{nm} 
\right ) g(\rho_\x,n),
\end{aligned}
\end{equation}
where $g(\rho_\x,n) = f(\rho_\x,n)/n! = \binom{\rho_\x}{n}$ is the 
number of (nonordered) $n$-tuples at site $\x$.
The first line of Eq. \eqref{Eq:ave} follows the usual convention
in the field theoretical study of boson systems and the second
line is introduced to save memories in actual simulations.
For a later purpose, we introduce a model dependent
function $h(\rho_\x,n)
= \epsilon_n g(\rho_\x,n)$, where
$\epsilon_n$ takes 1 (0) if
$D\delta_{n,1} + \sum_m \lambda_{n,m}$ is nonzero (zero).
The meaning of $\epsilon_n$ is straightforward; we  have only to consider
the dynamics with nonzero transition rate. 

The algorithm starts by selecting one of $n$-tuples at 
any site, randomly.
The simplest way to implement the selection is as follows:
First a site $\x$ is picked up with probability
$N_\x / M$, where $N_\x = {\sum_n} h(\rho_\x,n)$ is 
the number of accessible states (NAS) at site $\x$ and 
$M = \sum_\x N_\x$ is the total number of accessible
states (TNAS). 
Then, $n$ is chosen with probability $h(\rho_\x,n)/N_\x$. 
In this procedure, the array of the number of particles at all sites,
say $\brh[~]$ ($\brh[\x] = \rho_\x$), is  necessary.
However, it is not efficient as there are too many floating number 
calculations.  For a faster performance we introduce two more 
arrays, say \lis[~] and \act[~][~]. 
The array \lis[~] refers to the location of any $n$-tuple.
Each element of \lis[~] takes the form $(\x,\ell)$, 
where $\x$ is a site index and $\ell$ lies between 1
and the NAS at site $\x$.
From $\ell$ and the array $\brh[~]$, 
which $n$-tuple is referred to by the array \lis[~] is determined.
If $\ell \le h(\rho_\x,0)$, then $n=0$  is implied. 
Else if $\ell \le h(\rho_\x,0)+ h(\rho_\x,1)$, $n=1$ is meant. 
Else if $\ell \le h(\rho_\x,0)+h(\rho_\x,1) + h(\rho_\x,2)$, 
$\ell$ indicates one of pairs at site $\x$, and so on. 
In case the TNAS in the system is $M$, the size of \lis[~] is $M$ 
and all elements of \lis[~] should satisfy that 
$\lis[p] \neq \lis[q]$ if $p\neq q$ ($1\le p,q \le M$).
Hence, the random selection of an integer between 1 and $M$ 
is equivalent to choosing one of all $n$-tuples  with an equal probability. 
The array \act~is the inverse of the \lis. In other words, 
$\lis[s]=(\x,\ell)$ corresponds to $\act[\x][\ell]=s$.
It is clear that these two selecting mechanisms are equivalent
in the statistical sense.

After choosing $\x$ and $n$, the reaction $nA \rightarrow (n+ m)A$
occurs with  probability $n! \lambda_{nm} \Delta t$ for all
$m$, where $\Delta t$ is a configuration independent
time difference.  Provided $n=1$ is selected,
a particle at $\x$ hops to one of the
nearest neighbors with probability $D\Delta t$.
To make the transition probability meaningful, $\Delta t$ is made to satisfy
\begin{equation}
\left (2 d D \delta_{n,1} + \sum_m n! \lambda_{n,m} \right ) \Delta t \le 1,
\end{equation}
for all $n$. After this update, time increases by $\Delta t / M$.
On average, this algorithm generates $E(\Delta t,\{\rho\})$ 
transition events during $\Delta t$.

For systems with $k$ species, all we have to do is to 
modify the NAS at site $\x$ in such a way that
\begin{equation}
N_\x = \sum_{i=1}^k \sum_{n}h_i(\rho_{i,\x},n) + \sum_{n_1,\ldots,n_k}
h_{1,\ldots,k}(\rho_{1,\x}, \ldots,\rho_{k,\x};n_1,\ldots,n_k),
\label{Eq:multinas}
\end{equation}
where the first (second) terms are from the dynamics in which
$n$ particles of same species ($n_j$ particles of each $j$th species) 
are involved.
For instance, for the pair annihilation of different species, so-called
$A+B\rightarrow 0$ reaction, the second term of 
Eq. \eqref{Eq:multinas} becomes $\rho_{A,\x} \rho_{B,\x}$.
Except this modification, all other steps are the same as in the 
single species case.

Equipped with the numerical methods, Sec. \ref{Sec:app} studies
some bosonic RD systems which show scaling behavior.
\section{\label{Sec:app}Applications}
\subsection{\label{sec:AA}single species annihilation model}
The first example is the one dimensional 
single species annihilation model which corresponds to $\lambda_{nm}=0$
unless $n=2$ and $m=-2$.  For convenience, we set $D=\frac{1}{2}$
and $\lambda_{2,-2} = \lambda$.
The decaying behavior of the particle density was
studied in Ref. \cite{P05c}. This section  studies 
the correlation function $M(r;t)$ which is defined as
\begin{equation}
M(r;t) = \left \{
\begin{array}{lr}
\displaystyle
\lim_{L\rightarrow \infty} \frac{1}{L} \sum_{x=1}^L
\langle \rho_x(t) \rho_{x+r}(t) \rangle&\text{ if } r\neq 0,\\
\displaystyle
\lim_{L\rightarrow \infty} \frac{1}{L} \sum_{x=1}^L
\langle \rho_x(t) (\rho_x(t) - 1 ) \rangle&\text{ if } r = 0,
\end{array}
\right .
\end{equation}
where $\langle \ldots \rangle$ means the average over all independent 
realizations.
Using the boson operators in Ref. \cite{bosonF,Lee94,CT96}, 
$M(r;t)$ can be rewritten as $\frac{1}{L} \sum_x 
\langle a_x a_{x+r} \rangle$.

The correlation functions for the annihilation model of 
hard core particles with annihilation probability $p$ were
studied in Ref. \cite{PPK01}.
The asymptotic behavior of the correlation function 
is conjectured as \cite{PPK01}  
\begin{equation}
M_r(t) = \frac{1}{(4 \pi t )^{3/2}} \left ( \pi r + c \frac{1-p}{p} \right ),
\label{Eq:HC_DLAn_Mr}
\end{equation}
with $c=3.4\pm 0.2$. 
Note that 
$M_r(t)$ is not to be confused with $M(r;t)$; $M_r(t)$ and $M(r;t)$ are
defined in the hard core and boson models, respectively.

In fact, the exact value of $c$ can be deduced from the differential equation 
\begin{equation}
\frac{d \rho(t)}{dt} = - 2 p M_1(t),
\end{equation}
which relates the time 
derivative of the density $\rho(t)$ to the correlation function 
with $r=1$.
Since $\rho(t) \sim 1/\sqrt{4 \pi t}$ for any finite $p$ 
in the asymptotic regime,
it is easy to deduce that $c=\pi$ if the asymptotic behavior of the 
correlation function takes the form of Eq. \eqref{Eq:HC_DLAn_Mr}.
This value is compatible with the numerical estimation in Ref. \cite{PPK01}.

By the same token, we can conjecture how the
correlation function $M(r;t)$ behaves asymptotically from the equation
\begin{equation}
\frac{d \rho(t)}{dt} = - 2 \lambda M(0;t).
\end{equation}
If $M(r;t)$ takes the similar form to Eq. \eqref{Eq:HC_DLAn_Mr} and
since $\rho(t)$ decays as  $1/\sqrt{4\pi t}$ 
in the asymptotic regime for any nonzero value of $\lambda$ \cite{P05c}, 
one can deduce 
\begin{equation}
M(r;t)\sim M_\text{as}(r;t) \equiv
 \frac{\pi}{(4 \pi t)^{3/2}} \left ( r + \frac{1}{\lambda} \right)
\label{Eq:AA_Mr}
\end{equation}
for all $r \ge 0$. As far as we are aware of, 
the correlation functions of the boson annihilation model have not been studied
before. 
If $D \neq \frac{1}{2}$, the correlation function can be found 
by changing $t \mapsto 2 D t$ and $\lambda \mapsto \lambda/(2 D)$.
Since the boson model with infinite $\lambda$ is equivalent
to the hard core particle model with $p=1$ which is exactly
soluble, Eq. \eqref{Eq:AA_Mr} becomes exact in this limit; see Eq. 
\eqref{Eq:HC_DLAn_Mr}.

\begin{figure}[b]
\includegraphics[width=\textwidth]{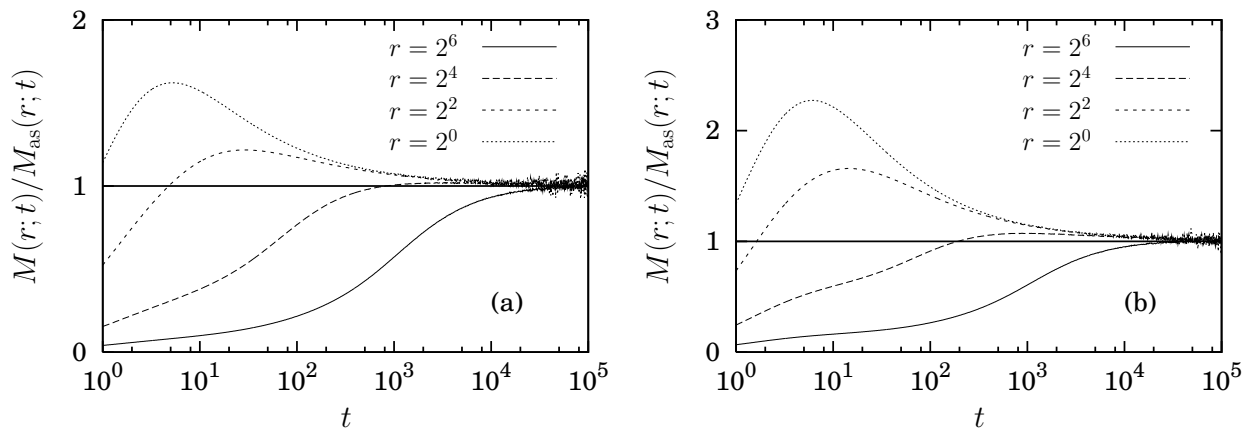}
\caption{\label{Fig:AA_Mr} 
Semi-log plots of $M(r;t)/M_\text{as}(r;t)$ as a function of $t$ 
for (a) $\lambda = 1$
and (b) $\lambda = \frac{1}{2}$.
All curves converge to 1 as $t$ goes to infinity.}
\end{figure}
In the following,
we will check the validity of  Eq. \eqref{Eq:AA_Mr} for 
finite $\lambda$ and nonzero $r$ via  MC simulations.
Initially, particles are distributed  according to
the uncorrelated Poisson distribution with average density $\rho_0=1$.
During simulations, 
we measured $M(r;t)$ for $r=2^0$, $2^2$, $2^4$, and $2^6$ up to
$t = 10^5$. The system size is $2^{18}$ and around $2.5\times 10^5$ 
independent samples are collected for both cases of 
$\lambda = 1$ and $\frac{1}{2}$.
Figure \ref{Fig:AA_Mr} shows that $M(r;t)$ takes the conjectured asymptotic
form \eqref{Eq:AA_Mr}.

The MC simulations of bosonic RD systems can confirm the 
equivalence between the (discrete) microscopic models and (continuous) 
Langevin equations, if exists.
From the coherent state path integral representation of the bosonic systems
\cite{bosonF},
Langevin equation can be derived in case each reaction
involves at most two particles.
Since the reaction of boson annihilation model requires two particles,
one can write down Langevin equation  which reads 
(It\^o interpretation is employed)
\begin{equation}
d a_x = dt(D \nabla^2_x a_x - 2 \lambda a_x^2) 
+ i \sqrt{2 \lambda} a_x dW_x ,
\label{Eq:AA_Langevin_eq}
\end{equation}
where $a_x$ is a complex stochastic random variable 
whose average is the mean number of particles at 
site $x$, $\nabla^2_x$ is the lattice Laplacian defined as
$\nabla^2_x f(x) = f(x+1) + f(x-1) - 2 f(x)$, $i$ is the imaginary number,
and $W_x$ is a Wiener process with 
$\langle dW_x dW_{x'} \rangle = dt \delta_{x,x'}$.
Initially, $a_x$ takes the value of $\rho_0$ 
which is the initial density of the uncorrelated Poisson distribution
used in the MC simulation.

\begin{figure}[b]
\includegraphics[width=0.5\textwidth]{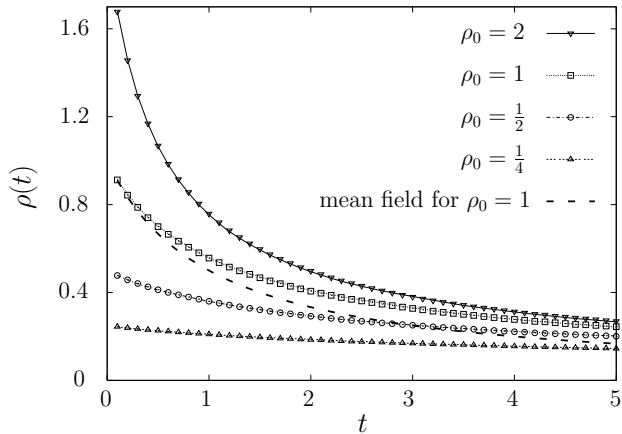}
\caption{\label{Fig:AA_MC_L} 
Plots of $\rho(t)$ obtained from MC simulations (lines) and
numerical integrations of Langevin equation (symbols) starting
from the initial density $\rho_0$.
The broken line without symbols
is the mean field solution of Eq. \eqref{Eq:AA_Langevin_eq}.
}
\end{figure}
This equation is integrated using Euler scheme with 
$\Delta t = 2.5\times 10^{-5}$ and the system size of  $2^{15}$. 
In Fig. \ref{Fig:AA_MC_L}, numerical integration
results for $\lambda = \frac{1}{2}$ are shown 
with comparison to MC simulations. Within statistical
error, these two approaches yield the same results.
Since the deviation from the mean field solution is evident, 
Langevin equation in the observation time properly appreciates 
the effect of noise.
Hence, we believe that Fig. \ref{Fig:AA_MC_L} shows the equivalence
of two approaches for the annihilation model.
Needless to say, the numerical integration of Langevin equation 
is a much harder job than the Monte Carlo simulation.

\subsection{cyclically coupled model}
The absorbing phase transition has been extensively
studied as a prototype of the nonequilibrium critical phenomena \cite{H00}.
The RG based on the boson systems has been applied successfully especially
to the directed percolation (DP) universality
class. Recently, the particle number probability distribution for
boson models belonging to the DP class was studied numerically  
and the RG prediction was confirmed again \cite{APP05}.

On the other hand, the pair contact process with diffusion (PCPD) 
defies any numerical and analytical conclusions to date \cite{HH04}.
Although the driven PCPD (DPCPD) studied in Ref. \cite{PhP05a} seems to 
conclude that the PCPD forms a different universality class from the DP,
recent extensive numerical study \cite{H05} 
revives the scenario that the PCPD will eventually be found to belong to
the DP class with a huge corrections to scaling.
Still, the universality classification for the one dimensional PCPD is 
yet to be settled unambiguously.

To make matters worse, the recent RG study shows that
the field theory starting from the single species master equation 
is not viable \cite{JvWOT04}, which was also anticipated independently
in Ref. \cite{PhP05a}. As both works conclude, the
field theory should account for the multispecies nature of the PCPD 
properly.  Following this instruction, multi component Langevin equations 
with real random variables are introduced and studied in Ref. \cite{DCM05}
to find a viable field theory for the PCPD.
We will take a slightly different path and ask 
whether we can find a viable field theory for the PCPD, in this section.

Since the PCPD involves two independent ``excitations'' such as
particles and pairs, it is natural to generalize to 
a two species model which captures the main physics of the PCPD.
This type of two species model with hard core particles was 
introduced and studied in Ref. \cite{H01}. This section introduces
a bosonic variant and studies it using both MC simulations and Langevin 
equations.

The model which will be called the cyclically coupled (CC) model 
is defined as follows:
There are two species, say $A$ and $B$. Each species diffuses with
rate $D_A$ and $D_B$, respectively.
Each $B$ particle is annihilated ($B\rightarrow 0$) with rate
$\delta$, branches another $B$ particle ($B \rightarrow 2 B$) with rate
$\sigma$, and mutates into two $A$ particles ($B \rightarrow 2 A$) with 
rate $\mu$. Every pair of $B$ particles at the same site can 
be coagulated ($2 B \rightarrow B$) with rate $\lambda$.
Every pair of $A$ particles produces a $B$ particle and is removed
($2 A \rightarrow B$) with rate $\tau$.
The $A$ ($B$) particles have a connection, if not a exact mapping, 
to the isolated particles (pairs) in the PCPD.

Since the PCPD as well as the CC suffers from the strong corrections to scaling,
it is nontrivial to show directly by MC simulations 
that the CC and the PCPD should belong to the same universality class.
Fortunately, we have an alternative to check the equivalence of the
CC and the PCPD in the sense of the universality.
If the relative bias between two species in the CC in one dimension 
triggers the mean field scaling with logarithmic corrections
as happens in the DPCPD \cite{PhP05a}, it is reasonable to conclude that
the CC and the PCPD share the critical behavior.

\begin{figure}[t]
\includegraphics[width=0.45\textwidth]{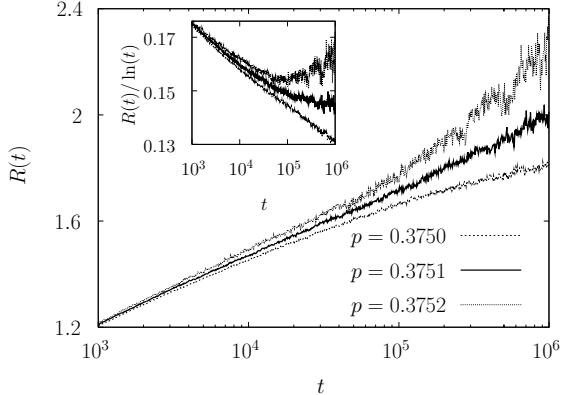}
\caption{\label{Fig:CC_bias} Semi-log plot of
$R(t)$ vs $t$ for the CC with the relative bias 
near criticality. At criticality ($p_c=0.3751$), 
clear logarithmic behavior is observed as in Ref. \cite{PhP05a}. Inset:
A plot of $R(t)/\ln(t)$ vs $t$ in the semi-log scales.}
\end{figure}

The transition events of the CC with a relative bias in one dimension are 
almost same as those of the CC above except that
$A$ particles hop only to the right with rate $1$.
For a numerical study, we set
$D_B = 0.1$,  $\mu = 0.2$, $\tau = 0.5$,
$\delta = 2 \lambda = 0.6\times p$, and $\sigma = 0.6 \times (1-p)$ with a
tuning parameter $p$.
Since only $A$ particles diffuse in a biased manner, the relative bias
between different species can not be gauged away by the Galilean transformation.
Figure \ref{Fig:CC_bias} shows that
$R(t)$ ($= A(t)/B(t)$) which is a ratio of two densities at time $t$
behaves logarithmically at criticality. 
Combining with the observation that
$A(t)\sim t^{-0.5}$ at criticality 
with possible logarithmic corrections (not shown),
the CC with the bias shows the same critical behavior as the DPCPD,
which confirms the equivalence of  the CC to the PCPD in the sense of
the universality. 
Accordingly, Langevin equations which are equivalent to the CC
are supposed to describe the critical behavior of the PCPD. 

Following standard path integral formalism \cite{bosonF}, one can derive
the action of the CC, which reads
\begin{equation}
\label{Eq:Action_CC}
\begin{aligned}
{\cal L} &=\bar a_x [ \partial_t a_x - D_A \nabla_x^2 a_x + v 
\partial_\| a_x - 2 \mu b_x + 2 \tau a_x^2]\\ &+\bar b_x [ \partial_t b_x - D_B \nabla_x^2 b_x 
- r  b_x + \lambda b_x^2 - \tau a_x^2 ]\\
& -\frac{1}{2} \bar b_x^2 ( 2 \sigma b_x - 2 \lambda b_x^2 ) 
- \frac{1}{2} \bar a_x^2 ( 2 \mu b_x - 2 \tau a_x^2 ),
\end{aligned}
\end{equation}
where the average of the field $a_x$ ($b_x$) corresponds to 
the density of species $A$ ($B$) at site $x$ and 
$r = \sigma - \mu - \delta$.
Along the parallel direction denoted as $\|$, 
$A$ particles hop to the right (left) with rate $D_A + v/2$ ($D_A - v/2$).
Since the number of barred fields  does not exceed two in
each term, one can write down the equivalent Langevin equations to
the action \eqref{Eq:Action_CC}, which read
\begin{subequations}
\begin{eqnarray}
d a_x &=& dt ( D_A \nabla_x^2 a_x - v \partial_\| a_x 
+ 2 \mu b_x - 2 \tau a_x^2 ) 
+ \sqrt{ 2 \mu b_x - 2 \tau a_x^2 } d W_x, \label{Eq:CC_A_eq}\\
d b_x &=& dt ( D_B \nabla_x^2 b_x + r b_x - \lambda b_x^2 + 
\tau a_x^2 ) 
+ \sqrt{2 \sigma b_x - 2 \lambda b_x^2} d V_x,
\end{eqnarray}
\label{Eq:CC_A}
\end{subequations}
where $W_x$ and $V_x$ are independent Wiener processes.

\begin{figure}[t]
\includegraphics[width=.95\textwidth]{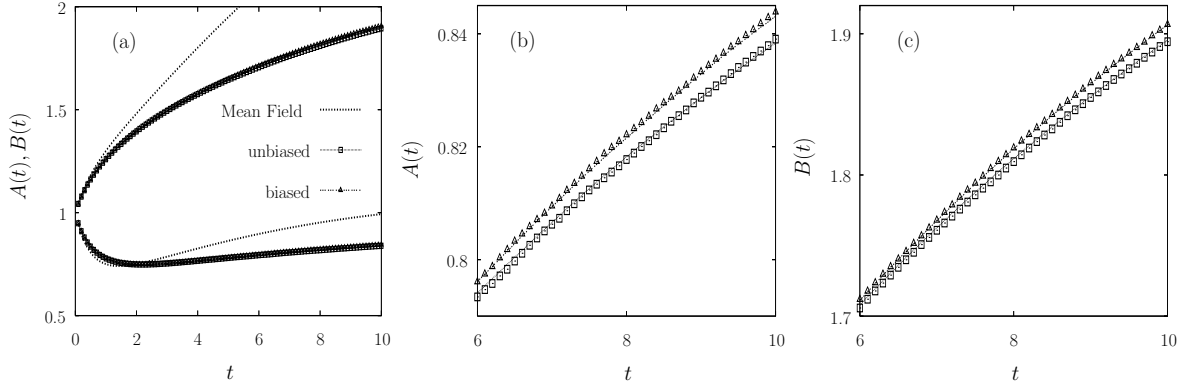}
\caption{\label{Fig:CCvsL} (a) The densities of each species
for both the biased and unbiased CC as a function of $t$
from the MC simulations (lines) and numerical integration (symbols)
of Langevin equations \eqref{Eq:CC_A}.
For comparison, mean field solutions are also shown. 
(b) Close-up of the interval $6 \le t \le 10$
for $A(t)$ in (a). (c) Close-up of the same interval as in (b), 
but the plots are for $B(t)$.}
\end{figure}
In Fig. \ref{Fig:CCvsL}, we compare the MC simulations of the CC with 
the numerical integrations of Langevin equations (\ref{Eq:CC_A}) 
at $p=0.29$. Initially, $a_x$ and $b_x$ are set to 1.
The system size for the numerical integration is $2^{15}$ and around
50 samples are independently generated with $\Delta t = 2.5\times 10^{-5}$.
Up to $t=10$, the difference between
the unbiased and biased cases is minute, but, within statistical errors, 
the behavior of two cases can be discerned from each other.
In other words, we showed that Eqs. \eqref{Eq:CC_A} are equivalent
to  the CC with/without bias. Although we compared two approaches just for
one set of parameter values, the full equivalence for all
parameter values is still expected. 

In summary, we showed that the CC and the PCPD 
share the critical behavior.  Then, we found Langevin equations 
which are equivalent to the CC. From these two observation,
we can say that Langevin equations \eqref{Eq:CC_A} with complex random variables
$a$ and $b$ show the same critical behavior as the PCPD.

Although we found the representative Langevin equations for the PCPD,
it is not obvious whether these equations with naive continuum limit
can serve as a properly coarse-grained field theory for the PCPD.
Besides, we are not sure whether Eqs. \eqref{Eq:CC_A} contain all relevant
(or sometimes dangerously irrelevant) terms. For example, the reaction
$A+B \rightarrow 0$ which is absent in our model can be
generated by a chain of reactions.
It is of no difficulty to write down Langevin equations with the 
pair annihilation of different species.
However, what will happen if we include the reaction $3A \rightarrow 0$
which prohibits writing down Langevin equations like Eqs. \eqref{Eq:CC_A}? 
If this reaction is also important in whichever sense (relevant or dangerously
irrelevant), terms with only $a$ and $\bar a$ in the
action take exactly the same form as those in Ref. \cite{JvWOT04}.
Hence, it seems that the difficulty found in Ref. \cite{JvWOT04} still remains
even in the multi component Langevin equations studied here.
We only hope that this study can be a starting point of the 
field theoretical understanding of the PCPD in the future.

\section{\label{Sec:sum} Summary}
To summarize, 
using the  algorithm  proposed in Ref. \cite{P05c} and
generalized one to the multispecies models,
the single species annihilation  and the cyclically
coupled models are studied.

For the single species annihilation model, 
the exact asymptotic form of the correlation functions
is conjectured  and numerically confirmed.
In addition, the equivalence of Langevin equation derived
from the coherent state path integral formalism 
to the discrete boson model is affirmed.
From the cyclically coupled model of bosons,
we derive Langevin equations for both biased and unbiased cases.
By simulating discrete models and integrating the Langevin equations 
numerically, these continuum equations are indirectly shown to 
describe the critical behavior of the PCPD and the DPCPD.

\end{document}